\documentstyle[aps,preprint,epsf,psfig]{revtex}

\newcommand{\ul}{\underline}
\newcommand{\kB}{k_{\mathrm{B}}}

\begin{document}

\date{\today}

\title{Domain State Model for Exchange Bias}

\author{U.\ Nowak, A.\ Misra, and K.\ D.\ Usadel} \address{Theoretische
  Tieftemperaturphysik, Gerhard-Mercator-Universit\"{a}t Duisburg,
  47048 Duisburg, Germany}

\maketitle

\begin{abstract}

  Monte Carlo simulations of a system consisting of a ferromagnetic
  layer exchange coupled to a diluted antiferromagnetic layer
  described by a classical spin model show a strong dependence of the
  exchange bias on the degree of dilution in agreement with recent
  experimental observations on Co/CoO bilayers. These simulations
  reveal that diluting the antiferromagnet leads to the formation of
  domains in the volume of the antiferromagnet carrying a remanent
  surplus magnetization which causes and controls exchange bias. To
  further support this domain state model for exchange bias we study
  in the present paper the dependence of the bias field on the
  thickness of the antiferromagnetic layer. It is shown that the bias
  field strongly increases with increasing film thickness and
  eventually goes over a maximum before it levels out for large
  thicknesses. These findings are in full agreement with experiments.
\end{abstract}
\pacs{75.70.Cn, 75.40.Mg, 75.50.Lk, 85.70-w}


For a ferromagnet (FM)  in contact with an antiferromagnet (AFM) a
shift of the hysteresis loop along the magnetic field axis can occur
which is called exchange bias (EB).  Usually, EB is observed after
cooling the system with the FM magnetized in saturation below the
N\'eel temperature $T_{\mathrm{N}}$ of the AFM or after cooling the
entire system in an external magnetic field. Although this effect is
well known since many years\cite{meiklejohnPR56,meiklejohnPR57} 
its microscopic origin is still discussed controversially. For a
review see a recent article by Nogu\'es and Schuller
\cite{noguesJMMM99}.

In a previous Letter \cite{miltenyiPRL00} we have reported on EB
observed experimentally in Co/CoO bilayers as a function of volume
defects in the antiferromagnet.  Of particular importance in this
study was the observation that it is possible to strongly influence EB
in Co/CoO bilayers by diluting the antiferromagnetic CoO layer, i.\ e.
by inserting nonmagnetic substitutions (Co$_{1-x}$Mg$_x$O) or defects
(Co\( _{1-y} \)O) not at the FM/AFM interface, but rather throughout
the volume part of the AFM. While the undiluted samples show only a
very small EB, dilution increases EB dramatically. Since for all samples
investigated a 0.4 nm thick CoO layer with minimum defect
concentration was placed at the interface the observed EB is primarily
not due to disorder or defects at the interface. Rather, the full
antiferromagnetic layer must be involved and we have argued that in
our systems EB has its origin in a domain state in the volume part of
the AFM which triggers the spin arrangement and thus the FM/AFM
exchange interaction at the interface. Indeed, in diluted
antiferromagnets when cooled in external fields metastable domains
occur carrying a surplus magnetization and having a very slow dynamics (for
reviews see \cite{kleemannIJMP93,belangerBOOK98}). These
domains are frozen to a large extend during hysteresis cycles and
their frozen magnetization is the origin of EB. This domain state model for EB
was supported by large scale Monte Carlo simulations performed at finite
temperatures. \cite{miltenyiPRL00,nowakPRB01}

To gain additional evidence for the domain state model in the present
paper we will concentrate on the dependence of EB on the thickness of
the AFM. The system consists of a FM monolayer exchange coupled to a
diluted AFM layer with $t$ monolayers.  The FM is described by a
classical Heisenberg model with vector spins ${\ul S}_i$ and exchange
constant $J_{\mathrm{FM}}$. The dipolar interaction is approximated by
an additional anisotropy term (anisotropy constant $d_x = -
0.1J_{\mathrm{FM}}$) which mimics the shape anisotropy leading to a
magnetization which is preferentially in the $y-z$-plane.  Also, we
introduce an easy axis in the FM ($z$-axis, anisotropy constant $d_z =
0.1J_{\mathrm{FM}}$) in order to obtain well defined hysteresis loops.
$d_z$ sets the Stoner-Wohlfarth limit of the coercive field, i. e. the
zero temperature limit for magnetization reversal by coherent rotation
($B_{\mathrm{SW}} = 2 d_z$, in our units, for a field parallel to the easy axis).
In view of the rather strong anisotropy in CoO we assume an Ising
Hamiltonian for the AFM. Thus the Hamiltonian of our system is given
by

\begin{eqnarray}
  {\cal H} = & - & J_{\mathrm{FM}} \sum\limits_{\langle i, j \rangle}
                 {\ul S}_i \cdot {\ul S}_j - \sum\limits_i
                 \left( d_z S_{iz}^2 + d_x S_{ix}^2
                 + {\ul S}_i \cdot {\ul B} \right)  \nonumber \\
            & - & J_{\mathrm{AFM}} \sum\limits_{\langle i, j \rangle}
                \epsilon_i \epsilon_j \sigma_i \sigma_j
               -\sum\limits_i B \epsilon_i \sigma_i \nonumber \\
            & - & J_{\mathrm{INT}} \sum\limits_{\langle i, j \rangle}
                          \epsilon_i \sigma_i S_{jz}.
\end{eqnarray}
with the antiferromagnetic nearest-neighbor exchange constant
$J_{\mathrm{AFM}} < 0$ and the effective in-plane magnetic field
$\ul{B} = B \hat{\ul{z}} + B_y \hat{\ul{y}}$.
The values of the magnetic moments are incorporated in $B$ and $d$,
respectively, so that the quantities ${\ul S}_i$ denote unit vectors
and $\sigma_i = \pm 1$ Ising spin variables. A fraction $p$ of the
sites of the lattice is left without a spin (quenched disorder:
$\epsilon_i = 0,1$).  For the exchange constant of the AFM which
mainly determines its N\'eel temperature (also depending on the
dilution, of course) we set $J_{\mathrm{AFM}} = - J_{\mathrm{FM}}/2$.
There seems to be some evidence that the exchange coupling between Co
and CoO is ferromagnetic \cite{miltenyiPRB01} but its strength is not
known experimentally. Therefore, we assume in our simulations a
ferromagnetic coupling with ($J_{\mathrm{INT}} = - J_{\mathrm{AFM}}$).

We use Monte Carlo methods with a heat-bath algorithm and single-spin
flip methods for the simulation of the model explained above. The
trial step of the spin update is a small variation around the initial
spin for the Heisenberg spins and --- as usual -- a spin flip for the
Ising spins \cite{nowakARCP00}. We perform typically 40000 Monte Carlo
steps per spin (MCS) for a complete hysteresis loop.
To observe the domain structure of the AFM we have to guarantee that
typical length scales of the domain structure fit into our system.
For the parameter values used in this simulation this is achieved for
systems of lateral extension $128 \times 128$.

In the simulations the system is cooled from above to below the
ordering temperature of the AFM in an applied external cooling field
$\ul{B} = B_c \hat{\ul{z}}$ with $B_c = 0.25J_{\mathrm{FM}}$ . The FM is then
long-range ordered and its magnetization is practically homogeneous
resulting in a nearly constant exchange field for the AFM monolayer at
the interface.  When the desired final temperature is reached a
magnetic field $\ul{B} = B \hat{\ul{z}} + B_y \hat{\ul{y}}$ is applied
which also has a small, constant perpendicular field component $B_y =
0.05J_{\mathrm{FM}}$ in order to define a certain path for the
rotation of the magnetization during field reversal and to avoid the
system to be trapped in a metastable state. The $z$ component of
the field $\ul{B}$ 
is then reduced in steps of $\Delta B = 0.004 J_{\mathrm{FM}}$ from
$B=0.2 J_{\mathrm{FM}}$ down to $-B$ and afterwards raised again up to the
initial value. 

Typical hysteresis loops are depicted in Fig.\ \ref{f:hysteresis-pk}.
Shown are results for the magnetization of the FM (upper figure) as
well as that of the AFM interface monolayer (lower figure) for
different thicknesses $t$ of the AFM. The hysteresis loops of the FM
clearly show EB depending on the thickness of the AFM. The
magnetization curves of the interface layer of the AFM are shifted
upwards due to the fact that after field cooling the AFM is in a
domain state with a surplus magnetization.  This layer experienced
during cooling the external field in addition to the exchange field of
the FM both having the same direction.  This shifted interface
magnetization of the AFM acts as an additional effective field on the
FM, resulting in EB. The hysteresis curve of the AFM interface layer
follows that of the FM layer with a much lower saturation
magnetization, however. With increasing thickness of the AFM layer the
area of the hysteresis loop of the interface layer which is
proportional to the energy losses in the AFM decreases indicating that
the spin structure in the AFM is stabilized.

The domain structure in the AFM interface layer is shown in Fig.\ 
\ref{f:domains} for an AFM consisting of one monolayer (upper figure)
and for ten monolayers (lower figure), respectively. The fractal
structure of these domains is obvious. It has been observed previously
in bulk systems and was analyzed in detail \cite{nowakPRB92,esserPRB97}.
The domain structures shown in Fig.\ \ref{f:domains} are to a large
extend frozen. But during field cycles small spin arrangements in the
domain boundaries can take place even at low temperatures resulting in
the hysteresis loops shown in Fig.\ \ref{f:hysteresis-pk}.

The structure of the domains depends on the thickness of the AFM. For
an AFM monolayer the effective field acting on all AFM spins is the
superposition of the strong exchange field and the external field. But
it is well known that the size of the domains depends on the strength
of the effective magnetic field. Large fields imply small domains and
vice versa
\cite{imryPRL75,birgeneauJSP84,belangerJAP85,nowakPRB92,esserPRB97}.
The small domains seen in the upper part of Fig.\ \ref{f:domains} are
thus due to the strong fields acting on the AFM monolayer. On the
other hand if the AFM consists of ten layers nine of them are only
exposed to the weak external field with the tendency to form larger
domain sizes. The coupling of these layers to the AFM interface layer
then results in a coarsening of the domains at the interface as seen
in the lower part of Fig.\ \ref{f:domains}. Note
that the distribution of vacancies in the interface layer is exactly
the same in both parts of Fig.\ \ref{f:domains}. A further obvious
consequence of this explanation is that  
the domain size becomes layer dependent
and increases with increasing distance from the AFM interface.
But after a certain distance from the interface the domain structure
should become independent of the interface layer which means that also
the bias field should become independent of the thickness of the AFM
for large $t$.  This behavior indeed is observed in our simulations.
In Fig.\ \ref{f:eb-vers-t} we show the dependence of the bias field on
$t$ for different dilutions of the AFM. The bias field is determined
as $B_{\mathrm{EB}} = (B^+ + B^-)/2$ where $B^+$ and $B^-$ are those
fields of the hysteresis loop branches for increasing and decreasing
field where the easy axis component of the magnetization of the FM
becomes zero. The absolute value of the bias field increases rapidly
with film thickness $t$, goes eventually over a maximum and then
levels out. This is in agreement with experiments
\cite{miltenyiPRB01}. Note that for the system with the smallest dilution the
absolute value of the bias field decreases for $t>1$ much
stronger with increasing thickness than for the more diluted films.
The reason is that the less diluted systems have a stronger tendency
to order antiferromagnetically thus reducing the net magnetization at
the AFM interface. Again, agreement with experiments is obtained.

In conclusion, we have found by Monte Carlo simulations further
support for our domain state model for exchange bias. So far all our
numerical results have been obtained for systems where the AFM has a
very strong anisotropy, i.e. behaves Ising-like. It is of great
interest to relax this condition, i.e. to consider vector spin models
for the AFM. Work in this respect is in progress.

{\bf Acknowledgments:} This work has been supported by the Deutsche
Forschungsgemeinschaft through SFB 491.

\bibliographystyle{myst}
\bibliography{Cite}

 
\begin{figure}
 \caption{Simulated hysteresis loops of the model explained in the
   text for $p = 0.4$ and $\kB T = 0.1J_{\mathrm{FM}}$. The field
   during cooling was $0.25 J_{\mathrm{FM}}$. Shown is the net
   magnetization in units of the saturation magnetization of the FM (upper
   figure) and of the interface monolayer of the AFM (lower figure)
   for different thicknesses $t$ of the AFM.}
 \label{f:hysteresis-pk}
\end{figure}

\begin{figure}[t]
  \caption{Frozen domain states of the AFM. Shown are staggered spin
    configurations (grey and black) of the AFM interface layer after
    the initial cooling procedure for dilution $p = 0.3$. AFM
    thickness $t=1$ (upper figure) and $t=10$ (lower figure).
    Vacancies are left white.}
 \label{f:domains}
\end{figure}

\begin{figure}
 \caption{Exchange bias field versus thickness $t$ of the AFM layer
    for different values of the dilution.}
    \label{f:eb-vers-t}
\end{figure}

\newpage
\centerline{\psfig{file=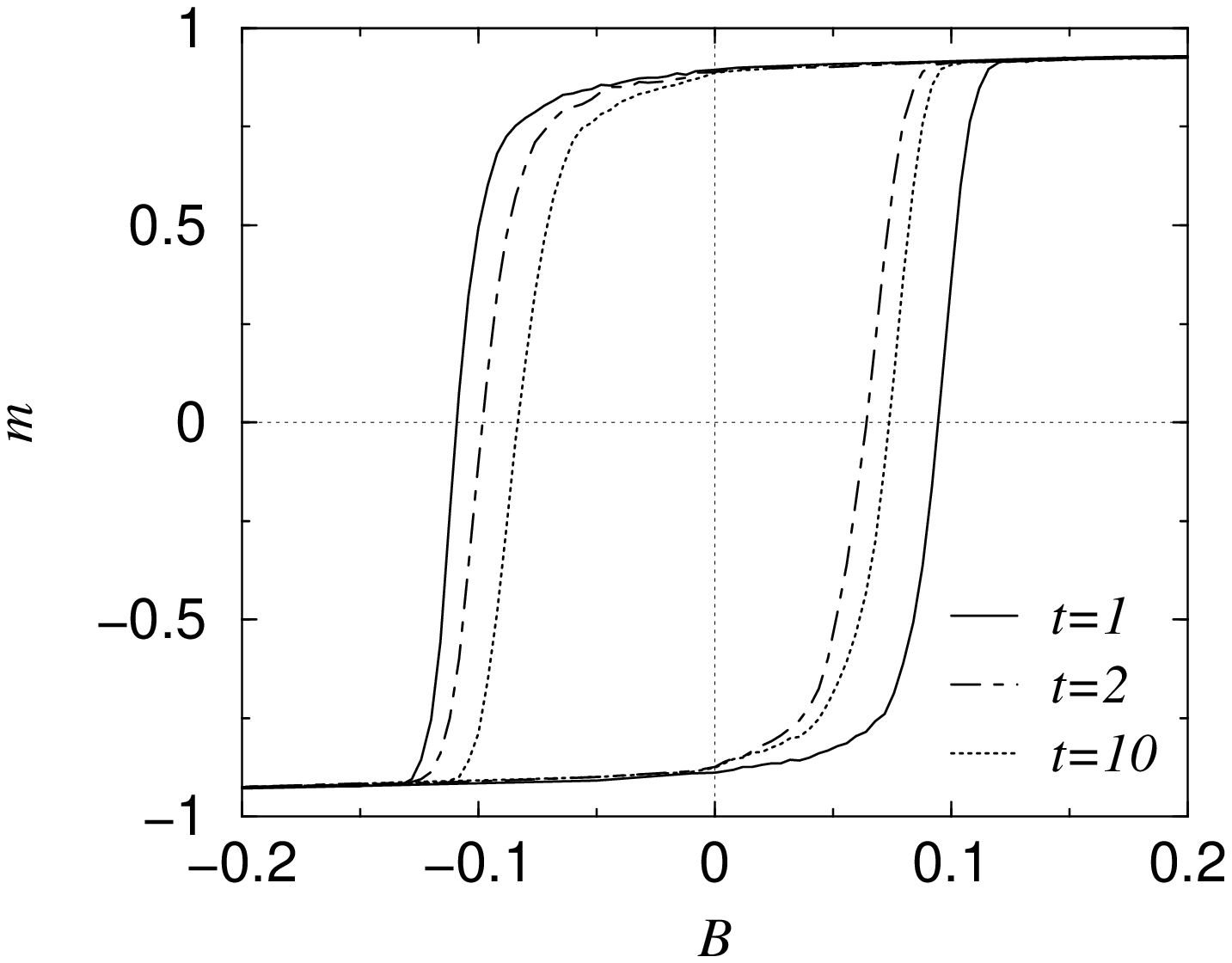,width=9cm,angle=0}}

 \centerline{\psfig{file=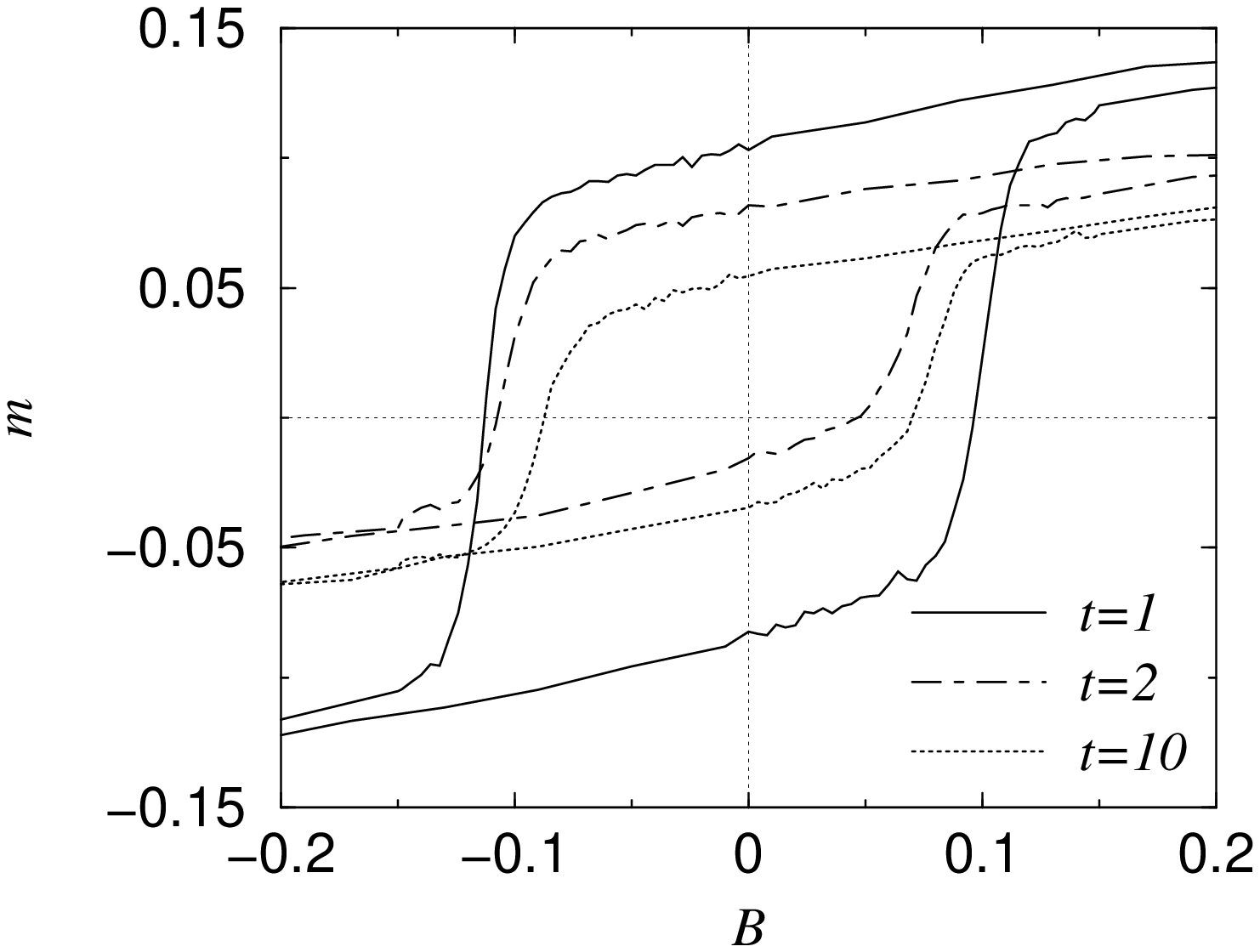,width=9cm,angle=0}}

 \vspace{3cm}
Figure 1: Nowak et al

\newpage
\epsfxsize=80mm
~ \hspace{2cm} \epsffile{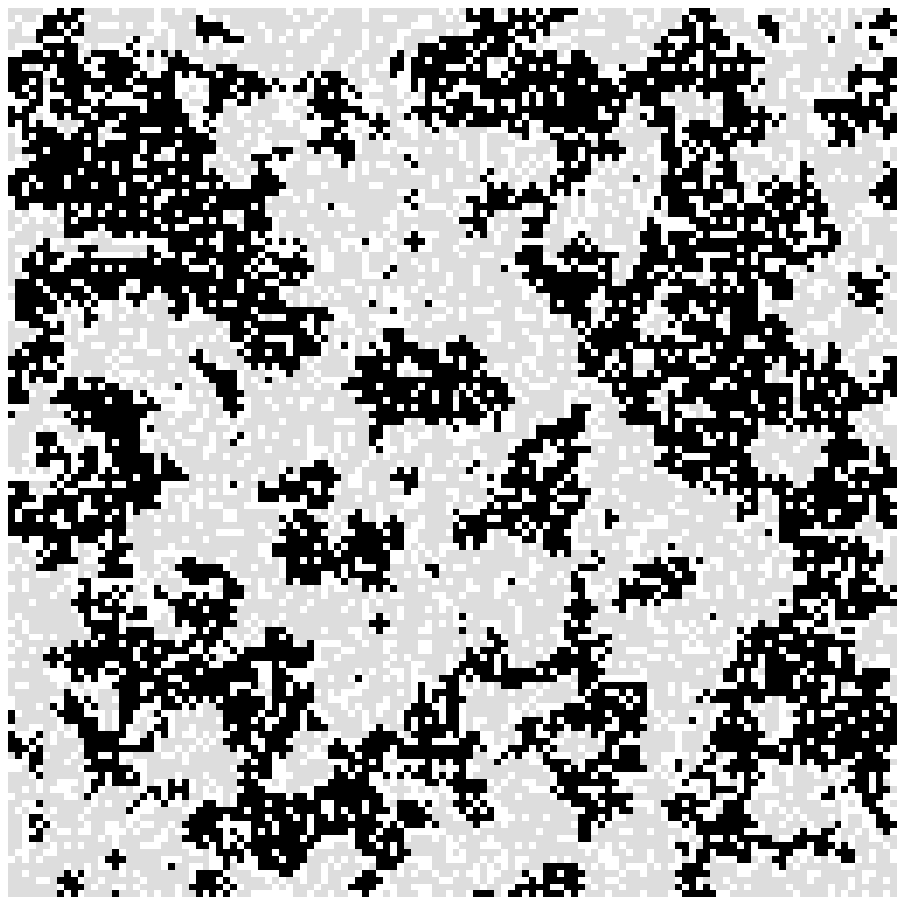}\\

\vspace{1cm}
\epsfxsize=80mm
~ \hspace{2cm} \epsffile{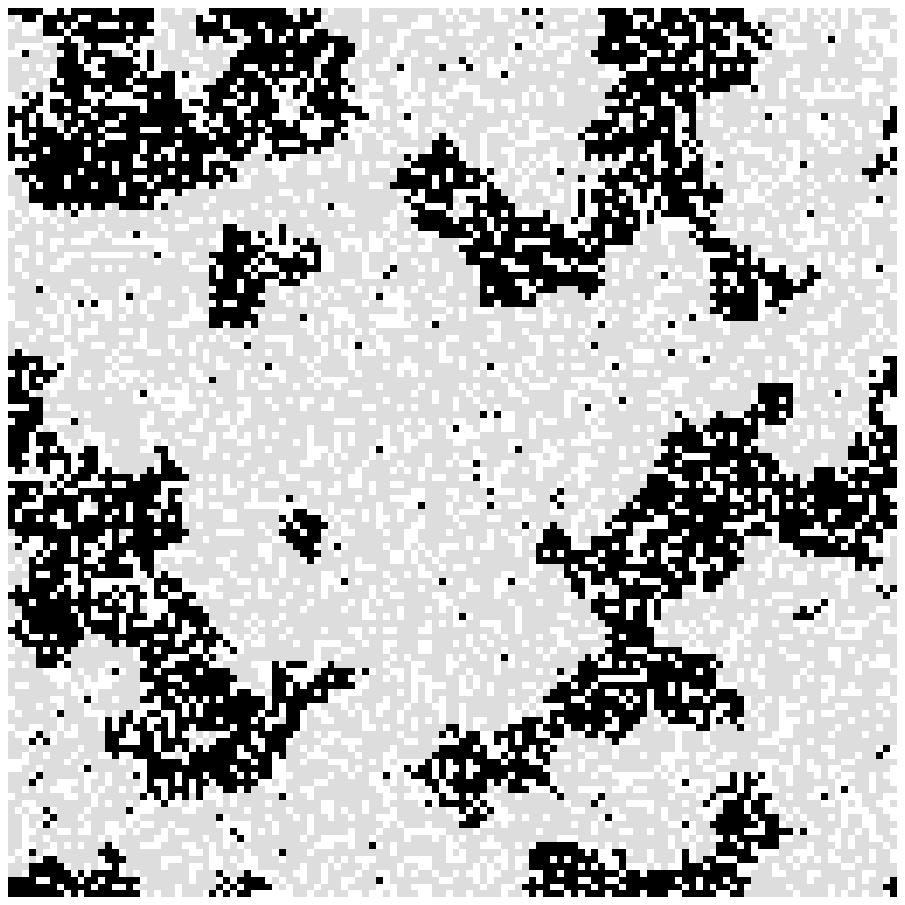}

\vspace{3cm}
Figure 2: Nowak et al
  
\newpage
 \centerline{\psfig{file=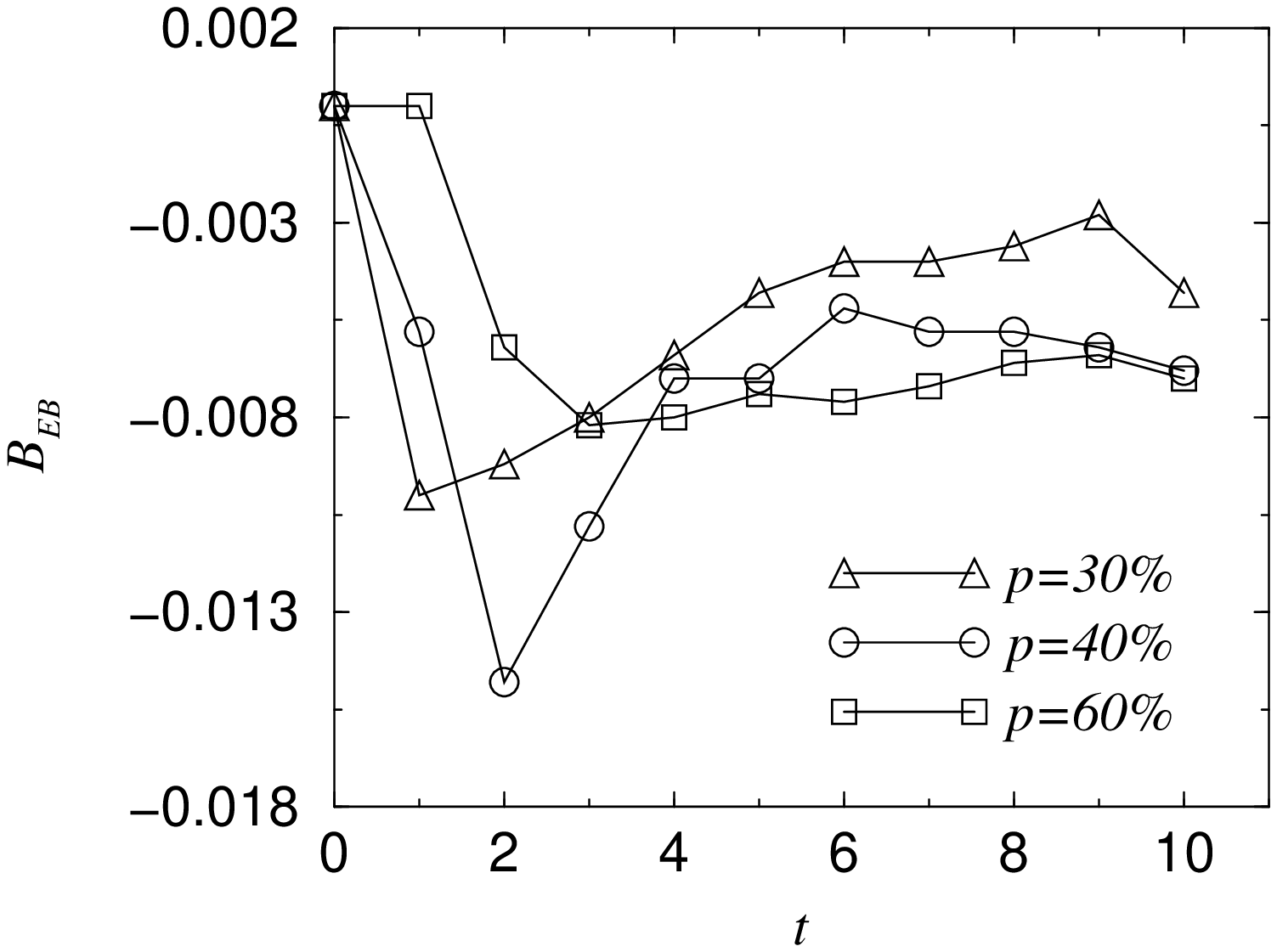,width=9cm,angle=0}}

\vspace{3cm}
Figure 3: Nowak et al

\end{document}